\documentclass{aa}
\usepackage{graphicx}
\usepackage{txfonts}
\usepackage{natbib}
\bibpunct{(}{)}{;}{a}{}{,}
\usepackage{multirow}

\begin{document}

\authorrunning{S\'odor et al.}

\title{The Blazhko behaviour of \object{\object{RR~Gem}inorum} II}

\subtitle{long-term photometric results}

\thanks{Tables $4-15$ are only available in electronic form at the CDS via anonymous ftp to cdsarc.u-strasbg.fr (130.79.128.5) or via http://cdsweb.u-strasbg.fr/cgi-bin/qcat?J/A+A/}

\author{\'A. S\'odor, B. Szeidl, \and J. Jurcsik}
\institute{Konkoly Observatory of the Hungarian Academy of Sciences. P.O.~Box~67, H-1525 Budapest, Hungary\\ \email{sodor,szeidl,jurcsik@konkoly.hu}}

\date{Received ; accepted }

\abstract
{\object{RR~Gem} is one of the few Blazhko RR~Lyrae that has photometric observations available extended enough to study the long-term courses of its pulsation and modulation properties in detail.}
{We investigate the pulsation and modulation properties and the relations between them in \object{RR~Gem} using photometric observations from the past 70 years in order to gain further insight into the nature of the Blazhko modulation.}
{We studied the photographic, photoelectric, and CCD light curves obtained at the Konkoly Observatory and other authors' published maxima observations. Detailed analysis of the light curves, maximum brightness, and $O-C$ data are carried out.}
{\object{RR~Gem} showed modulation most of the time it was observed. The modulation amplitude showed strong variations from the undetectable level (less than 0.04\,mag in maximum brightness) to about 0.20\,mag. The amplitudes of the amplitude and phase modulations showed parallel changes, thus the total `power' of the modulation have changed during the past 70 years. Parallel changes in the pulsation and modulation periods occur with a $\mathrm d\,P_\mathrm{mod}/\mathrm d\,P_\mathrm {puls}=1.6\pm0.8\times10^3$ ratio. We also detected $0.05-0.1$\,mag changes in the mean maximum brightness and mean pulsation amplitude.}
{}
\maketitle

\section{Introduction}

There is no satisfactory and commonly accepted explanation yet for the phenomenon of the light curve modulation of RR Lyrae stars, the century's old Blazhko effect \citep{blazhko}. Less than a dozen Blazhko RR Lyrae have photometric observations available that are extended enough to study their long-term modulation properties in detail. \object{RR~Gem} (\hbox{$\alpha=\mathrm{07^h 21^m 33\fs53}$}, \hbox{$\delta=+30\degr 52\arcmin 59\farcs5$}, J2000) is one of these stars that has been observed at the Konkoly Observatory from time to time since 1935. The observations were always obtained with the most up-to-date photometric technique available at the time: between 1935 and 1953 photographic, from 1954 to 1983 photoelectric, and in 2004 and 2005 CCD observations. (The photographic, photoelectric, and the 2005 CCD observations have not been published yet.)

\object{RR~Gem} was claimed to be Blazhko-modulated with a modulation period of 48\,days based on the photographic observations taken in the 1930s at the Konkoly Observatory \citep{balazs}. Although the modulation period was dubious, the fact of modulation itself seemed to be unambiguous. In contrast, in the 1950s the modulation seemed to cease \citep{detre}. We continued to observe \object{RR~Gem} in 2004 and 2005 and found it to be modulated but with very low amplitude and with the shortest period (7.216\,d) known at that time \citep{j05,padeu}.

In the first part of this series of papers, the results of the 2004 CCD observations of \object{RR~Gem} were reported \citep[ Paper1 hereafter]{j05}. This second part reports our results of the reexamination of all the available photometric data of \object{RR~Gem} in order to reveal the long-term behaviour of the modulation.

\section{Data}

We utilise photometric observations made by different instruments of the Konkoly Observatory. Most of these data are previously unpublished archive observations. We also use maxima timing observations of the past 114 years from the literature.

\begin{table*}[hhhttt!!!]
  \centering
  \caption{Log of the Konkoly observations of \object{RR~Gem}.}
  \label{tbl_lcsum}
  \begin{tabular}{clclccccrcc}
    \hline\hline
    \noalign{\smallskip}
\multicolumn{2}{c}{begin}  & \multicolumn{2}{c}{end} & telescope\,$^a$ & detector & filters & time res. & \multicolumn{1}{c}{data} & \multicolumn{1}{c}{observed} & comp. \\
{[JD]} & \multicolumn{1}{c}{date} & {[JD]} & \multicolumn{1}{c}{date}   &               &          &          &      [min]\,$^b$ & \multicolumn{1}{c}{points} & nights/maxima & star\,$^c$ \\
    \noalign{\smallskip}
    \hline
    \noalign{\smallskip}
2\,427\,833 & 1935 Jan & 2\,434\,457 & 1953 Mar & 16\,cm & pg & unf. & $4-6$ & 668 & 37/19 &  \\
2\,435\,052 & 1954 Nov & 2\,436\,227 & 1958 Jan & 60\,cm & pe & unf. & $1-4$ & 604 & 22/19 & A \\
2\,436\,229 & 1958 Jan & 2\,443\,931 & 1979 Feb & 60\,cm & pe & $BV$ & $2-8$ & 1\,754 & 37/28 & A \\
2\,441\,679 & 1972 Dec & 2\,445\,673 & 1983 Dec & 50\,cm & pe & $UBV$ & $2-8$ & 440 & 4/4 & A \\
2\,453\,019 & 2004 Jan & 2\,453\,440 & 2005 Mar & 60\,cm & CCD & $BV(RI)_\mathrm C$ & $4-8$ & 13\,355 & 63/31 & B \\
  \noalign{\smallskip}
  \hline
  \noalign{\smallskip}
  \multicolumn{10}{p{153mm}}{$^a$ See details about the telescopes in Sections~$\ref{sect_pgdata} - \ref{sect_ccddata}$.}\\
  \multicolumn{10}{p{153mm}}{$^b$ Time resolution means the typical cycle times in the case of multicolour observations.} \\
  \multicolumn{10}{p{153mm}}{$^c$ Comparison stars used are: A - \object{GSC 02452-01847}; B - \object{BD~$+31\degr1547$}\,\footnotemark[2].}\\
    
  \end{tabular}
\end{table*}

\subsection{Photometric data}

The light curves we use were obtained at the Konkoly Observatory between 1935 and 2005. Only a limited number of photoelectric data have been published by other authors \citep{fitch,epstein,stepien,liuj,esa}. These data are, however, few in number and too sparse for studying the modulation behaviour.

The log of the Konkoly observations is found in Table~\ref{tbl_lcsum}. The photographic, photoelectric, and 2005 CCD data, listed in \hbox{Tables~$4-12$}, are only available electronically at the CDS (http://cdsweb.u-strasbg.fr/cgi-bin/qcat?J/A+A/). In each table, column 1 lists the HJD of the observations. In Table~4, column~2 gives the photographic $B_\mathrm{pg}$ magnitudes, while in \hbox{Tables~$5-12$} differential $U, B, V, R_\mathrm c, I_\mathrm c$ magnitudes of \object{RR~Gem} are given with respect to \object{GSC 02452-01847} and \object{BD~$+31\degr1547$}\,\footnotemark[2] for the photoelectric and CCD data, respectively.

\subsubsection{Photographic data}
\label{sect_pgdata}

The photographic observations were made with the 16\,cm astrograph $(f = 2240\,\mathrm{mm})$ of the Konkoly Observatory at Budapest, Sv\'abhegy. Between 1935 and 1953, about 1000 exposures were taken on about 80 photographic plates. Unfortunately, many of the plates were lost or damaged during the past 70 years. We succeeded in recovering 598 measurable exposures on 56 plates.

The photographic plates were digitized on a Umax PowerLook 3000 flatbed transparency scanner with a spatial resolution corresponding to about 1.8\,"/pixel. Digital aperture photometry was applied to the images using standard IRAF\footnote{{\sc IRAF} is distributed by the National Optical Astronomy Observatories, which are operated by the Association of Universities for Research in Astronomy, Inc., under cooperative agreement with the National Science Foundation.} packages. In this way, we determined the photographic blackenings of \object{RR~Gem} and of the photographic comparison stars.

A series of comparison stars were chosen from the 20\,arcmin vicinity of \object{RR~Gem} to cover its brightness variation range. Altogether, 15 appropriate comparison stars were found. We measured the instrumental $B$ magnitudes of the comparison stars during the course of the 2005 CCD observations. The differential magnitudes were determined with respect to the $B$ magnitude of \object{BD~$+31\degr1547$}\,\footnote{This star was used as comparison for the $2004-2005$ CCD observations. In the first paper of this series, the wrong BD number was erroneously given (BD~$+31\degr1549$) for this star.} ($B = 10.643\,\mathrm{mag}$) we had derived earlier (see Paper1). The $B_\mathrm{pg}$ magnitudes of \object{RR~Gem} were determined by a $3^\mathrm{rd}$ order polynomial blackening curve fit for the comparison stars of each exposure. The most deviating points were omitted from the fits.

To check the consistency between our instrumental CCD $B$ photometry and the photographic measurements, we also automatically computed the magnitudes of the comparison stars for each of the exposures by the aforementioned blackening curve-fitting process using all the other comparison stars except for the most deviating ones. The magnitudes determined in this way showed good agreement with the accepted instrumental CCD $B$ magnitudes for all the 15 comparison stars. The differences have a mean of 0.00\,mag with a standard deviation of 0.05\,mag. Neither colour nor brightness dependency of the residuals was found according to our CCD measurements.

A record of the evaluation of only a part of the photographic observations has remained available. It contains 331 photometric measurements of \object{RR~Gem}, 70 of them from damaged or lost plates. These 70 points were brought onto our magnitude scale with a transformation that was derived from a linear regression of the 261 common data points. The photographic light curve was complemented with these 70 measurements.

\subsubsection{Photoelectric data}

The photoelectric observations were made with different unrefrigerated photometers between 1954 and 1983. The photometers in the Newton focus of the 60\,cm Heyde telescope $(f = 3600\,\mathrm{mm})$ at Budapest, Sv\'abhegy, first employed an RCA~1P21 and later an EMI~9052~B photomultiplier tube. In the Cassegrain focus of the 50\,cm telescope $(f = 7500\,\mathrm{mm})$ at the Piszk\'estet\H o Mountain Station, a photometer was used with an EMI 9058 QB multiplier. All the instruments were equipped with conventional Schott filters ($U$ - UG2~2mm, $B$ - BG12~1mm\,+\,GG13~2mm, $V$ - GG11~2mm), except for the first observations, which were made without filter. A standard photoelectric reduction procedure was applied. No correction was applied to the differential extinction, as the comparison star is as near as 8' to the variable. The colour extinction was also neglected because the $B-V$ colour of the variable and the comparison match at maximum brightness. The filtered measurements were transformed to the standard Johnson $UBV$ system according to the regularly determined telescope constants from standard star measurements. Measurements in the $U$ band were only made on 3 nights in the past few years of the photoelectric observations. These data are also published electronically but not used in the analysis. For the photoelectric observations, the star GSC~02452-01847 was used as comparison. We derived the standard $B$ magnitude of $10.79 \pm 0.01\,\mathrm{mag}$ of this star from the $2004-2005$ CCD measurements.

\begin{figure*}[]
  \centering
  \includegraphics[width=18cm,height=7.2cm]{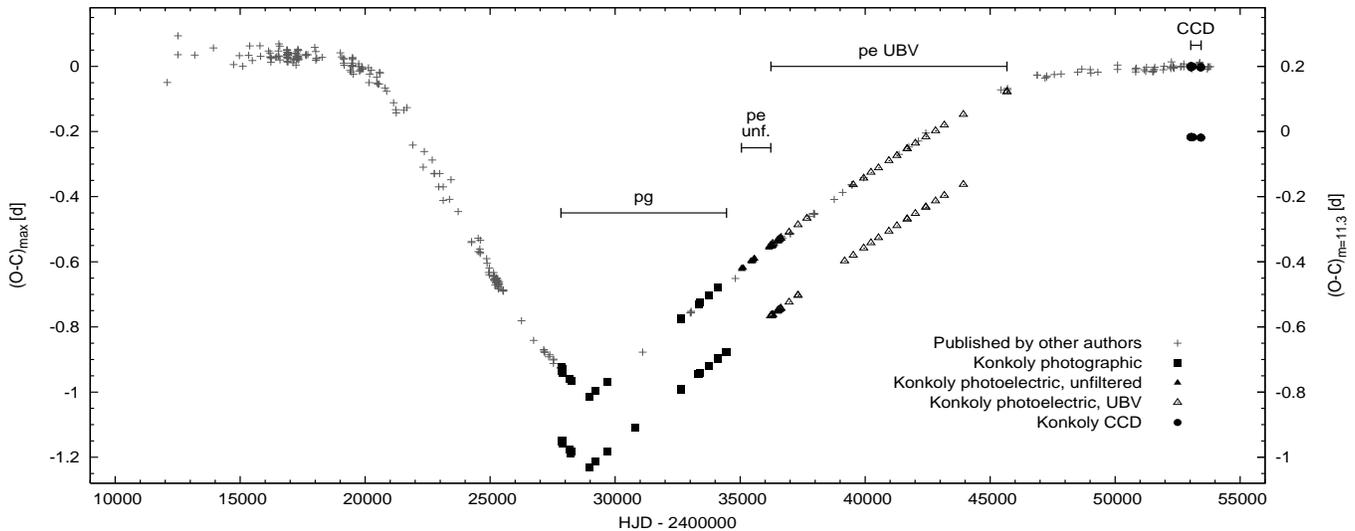}
  \caption{The $(O-C)_\mathrm{max}$ diagram of \object{RR~Gem} over the past 115 years according to the average pulsation period (upper line of points). The lower line of points represents the $(O-C)_{\mathrm m=11.3}$ data of the Konkoly light curves, calculated using the same period. These points are shifted downwards by 0.2\,d for better visibility (see the axis on the right side). The intervals of Konkoly observations are indicated.
  }
  \label{fig_oc}
\end{figure*}

\subsubsection{CCD data}
\label{sect_ccddata}

The 2004 CCD observations and results have been published in Paper1. In order to refine the pulsation and modulation periods and to measure $B$ magnitudes of stars in a larger area to calibrate the photographic data, further CCD observations were obtained in 2005 \citep{padeu} with the same instrumentation (the 60\,cm Heyde telescope with a 1152\,x\,770 Wright CCD detector in its Newton focus, which gives a field of view of 24'\,x\,17'). These data were reduced and transformed to the standard system in the same way as for the 2004 observations (see Paper1 for details).

\subsection{The $O-C$ data}

The changes in pulsation period and the phase modulation properties are studied using the $O-C$ values of the maximum brightness times and the times of the \hbox{$B=11.3\,\mathrm{mag}$} brightness on the rising branch, referred to as $(O-C)_\mathrm{max}$ and $ (O-C)_{\mathrm{m}=11.3}$, respectively. The $t_\mathrm{max}$ and $t_{\mathrm{m}=11.3}$ timings of the maxima and rising branches were determined from the Konkoly observations. The $O-C$ values were calculated according to the corresponding ephemerides, which are given in the subsequent sections.

\subsubsection{$(O-C)_\mathrm{max}$}
\label{sect_ocmax}

Most of the published times of maxima were collected in the GEOS\footnote{\tt http://dbrr.ast.obs-mip.fr/} database \citep{graff,waterfield,luizet,ivanova,ivanovb,chudovichev,dubiago,blazhko2,lange1,lange2,kleissen,batyrev1,batyrev2,alania,guriev,mandel,fitch,epstein,tsesevich,kl,langeetal,braune,liuj,esa,vdb1,vdb2,vdb3,vdb4,vdb5,vdb6,gensler,ah,ah2,ah3,hubs,hubs2,hubs3,agetal,hubsetal,leborgne,lb1,lb2,wils}. A few additional maxima times of \cite{waterfield} and \cite{graff} were also considered. We have left out some outlying visual points where no cause of the deviation was found. Altogether, 289 reliable $(O-C)_\mathrm{max}$ data points were used. Further  101 $(O-C)_\mathrm{max}$ points were determined from the Konkoly light curves by $3^\mathrm{rd}-5^\mathrm{th}$ order polynomial fits to the data points around the maxima. The times of the maxima were averaged for the $B$, $V$, $R_\mathrm c$, and $I_\mathrm c$ bands in the case of multicolour observations, since the times of maximum brightness do not differ significantly in these bands. Times of maxima in the $U$ band were not taken into account because of the larger uncertainty of these data. The fits also yielded the magnitudes of maximum brightnesses, which were utilised in studying the modulation behaviour.

\subsubsection{$(O-C)_{\mathrm{m}=11.3}$}

The phase modulation can be studied through the $O-C$ data of a particular point of the rising branch or of the maximum brightness. Because the time of maximum brightness can be determined with less accuracy than the time of a certain magnitude phase on the rising branch, we decided to investigate the modulation of the 11.3\,mag brightness phase on the rising branch in the $B$ band. This brightness approximately bisects the range from the middle of the rising branch to the mean maximum light.

The $(O-C)_{\mathrm{m}=11.3}$ variation measures, in fact, a combination of the amplitude and phase modulations. It shows oscillations due to the changing slope of the rising branch when the light curve change can be described as only a scaling in magnitude (exact amplitude modulation), as well as due to exact phase modulation. Just to estimate the contribution of the exact amplitude modulation in the amplitude of the $(O-C)_{\mathrm{m}=11.3}$, test data were generated using the mean CCD $B$ light curve of \object{RR~Gem} and assuming exact amplitude modulation. The modulation properties of the test data are compared with the observations of \object{RR~Gem} in Sect.~\ref{sect_nature}.

The phase modulation can be studied in the $(O-C)_{\mathrm{m}=11.3}$ data only if the effect of the long-term period changes are eliminated. Therefore we used different ephemerides for the different parts of the observations, as the period changes require. The ephemerides applied to construct the $(O-C)_{\mathrm{m}=11.3}$ data for the different parts of the observations are given in the corresponding sections.
\\

The maximum brightness times, magnitudes, and $t_{\mathrm{m}=11.3}$ data are given in \hbox{Tables~$13-15$} for the photographic, photoelectric, and CCD observations, respectively. These tables are only available electronically at the CDS (http://cdsweb.u-strasbg.fr/cgi-bin/qcat?J/A+A/). In all 3 Tables, columns~$1-3$ give the $t_{\mathrm{m}=11.3}$ data, their errors, and the times of the maximum light, respectively. In Table~13, columns~4 and 5 contain the $B_\mathrm{pg\,max}$ magnitudes and their errors, respectively. In Table~14, columns~4 and 6 list the photoelectric $\Delta B_\mathrm{max}$ and $\Delta V_\mathrm{max}$ magnitudes, and columns~5 and 7 give their errors, respectively. In Table~15, columns~4 and 5 list the CCD $\Delta B_\mathrm{max}$ magnitudes and their errors.

\section{The $O-C$ diagram}

The variation in the pulsation period during the past 115 years can be followed in the $O-C$ diagrams shown in Fig.~\ref{fig_oc}. The diagrams were constructed using the ephemerides 
$$ t_{\rm max} = 2412077.521\,{\rm [HJD]} + 0\fd397291066 \cdot E, $$
$$ t_{\rm m=11.3} = 2412077.501\,{\rm [HJD]} + 0\fd397291066 \cdot E. $$
The period corresponds to the average pulsation period of \object{RR~Gem} over the past 115 years. The $(O-C)_{\mathrm{m}=11.3}$ data calculated with this single ephemeris shows similar changes to the $(O-C)_\mathrm{max}$ data.

The $(O-C)_\mathrm{max}$ diagram indicates steady and abrupt pulsation period changes. The period of the pulsation varied between the extrema of 0.397253\,d and 0.3973177\,d. The transition between these values occurred abruptly just in the middle of the photographic observations, around JD\,2\,428\,800. After the period jump, the period was decreasing during the next 20000 days \citep[see Fig.~1 in][]{padeu} and seem to remain stable in the past ten years. Details of the period change rates and observed period values are given in the next sections.

Note that the shape of the $(O-C)_\mathrm{max}$ diagram, in particular the aforementioned sudden period change, strongly resembles that of the Blazhko RR Lyrae stars, \object{XZ~Cyg} \citep{aavso} and \object{RW~Dra} \citep{balazsdetre}.

\section{Light curve analysis}

The Konkoly photometric data were analysed in order to detect the modulation if existed, to follow any changes in the pulsation and modulation characteristics, and to find possible connections between them.

The observations were divided into several parts whenever the different observing methods, the data distribution in time, and the changes in the pulsation period allowed. When the phase coverage and data quality made it possible, Fourier analysis was performed to search for modulation components in the spectra. The modulation was also examined through the maximum brightness and the $(O-C)_{\mathrm{m}=11.3}$ data points calculated using the actual mean pulsation values. In this way, the maximum brightness variation caused by amplitude modulation and the phase modulation of the rising branch were studied.

Data analysis was performed using the different applications of the MUFRAN package \citep{mufran} and the linear and nonlinear curve fitting abilities of gnuplot\footnote{\tt http://www.gnuplot.info/} and Mathematica\footnote{Mathematica is a registered trademark of Wolfram Research Inc.} \citep{wolfram}.

\subsection{Results from photographic observations}
\label{sect_pglc}

Because of the sudden and remarkable period change that took place at the middle of the photographic observations, this data set was divided into two parts separated at JD\,2\,428\,800 (denoted as pg1 and pg2, respectively). In this way, we can follow whether the physical processes that gave rise to the sudden and significant change in the pulsation period have any effect on the modulation properties.

The pg2 data set has a much less favourable distribution in time than that of the pg1 data. There are 9 light maxima and 9 rising branches from two consecutive seasons in pg1, whereas we have 10 observed maxima and 15 rising branches from the 15-year long pg2 interval.

\begin{figure}[]
  \centering
  \includegraphics[width=8cm,height=9cm]{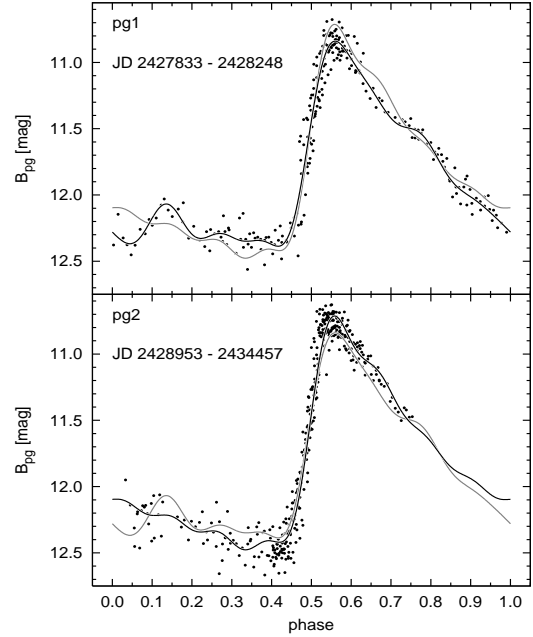}
  \caption{Konkoly photographic light curves for the pg1 and pg2 intervals. Solid lines show the mean light curves that  take 8 harmonics into account. Grey lines show the mean curve of the other data set to help compare the pulsation amplitudes. The times of the plotted pg2 light curve were transformed in order to eliminate the period change.}
  \label{fig_pg_lc}
\end{figure}

The pulsation period was determined by Fourier fit to the light curves taking 8 harmonics into account. The pg2 data set does not cover each phase of the pulsation, since the observations were focused on the rising branch and maximum light. The gap in the middle of the descending branch was bypassed with artificial points based on the CCD $B$ mean light curve, shifted and scaled appropriately. These points were used only to stabilize the Fourier fit so they had no effect on any of the conclusions. The $t_{\mathrm{m}=11.3}$ data show that a slight period decrease occurred during the pg2 interval, therefore a linear period change was taken into account. The pulsation periods derived for the two parts of the photographic data are
$$P_\mathrm{pg1} = 0.397253 \pm 5 \cdot 10^{-6}\,\mathrm{d},$$
$$P_\mathrm{pg2} = P_\mathrm{pg2\,0} + \dot{P}_\mathrm{pg2} \cdot (t-t_\mathrm{pg2\,0})\,\mathrm{d},$$
where $$P_\mathrm{pg2\,0} = 0.3973177\pm 2 \cdot 10^{-6}\,\mathrm{d},$$
$$\dot{P}_\mathrm{pg2} = -(3.7 \pm 0.8) \cdot 10^{-10}\,\mathrm{d/d},\ \mathrm{and}\ t_\mathrm{pg2\,0} = 2\,428\,953.0\,\mathrm{[HJD]}.$$
The light curves of the two intervals folded with the appropriate pulsation periods are plotted in Fig.~\ref{fig_pg_lc}. The fitted mean light curves are also shown in the plots. The plotted pg2 light curve was transformed to constant period by a HJD transformation \citep[see details in][]{omegacen}.

From the $t_{\mathrm{m}=11.3}$ times (listed in Table~13), $(O-C)_{\mathrm{m}=11.3}$ values were calculated according to the following ephemerides:
$$ t_{\mathrm{m=11.3\,pg1}} = 2\,427\,854.9870\,{\rm [HJD]} + P_{\mathrm{pg1}} \cdot E_{\mathrm{pg1}}, $$
$$ t_{\mathrm{m=11.3\,pg2}} = 2\,428\,953.0176\,{\rm [HJD]} + P_{\mathrm{pg2\,0}} \cdot E_{\mathrm{pg2}} - 7.3 \cdot 10^{-11} \cdot E_{\mathrm{pg2}}^2. $$
\noindent for the two intervals, respectively.

The light curves shown in Fig.~\ref{fig_pg_lc} indicate that the mean pulsation amplitude had increased for the second interval perceptibly. No differences in the plate material and in the data evaluation of the pg1 and pg2 observations account for these discrepancies. 

\subsubsection{Modulation properties of the Pg1 data (JD\,$2\,427\,833-2\,428\,248$)}

The rising branch of the pg1 light curve shows greater scatter around the mean curve than that of the pg2 data. Possible explanations of this phenomenon are incessant period change or phase modulation during the pg1 interval. Because the $(O-C)_{\mathrm{m}=11.3}$ values change sign more than twice during the pg1 interval, these changes cannot be explained with a unidirectional period change. Recently \cite{cc} have reported an irregular behaviour of the Blazhko modulation of \object{RR~Lyr}, and they explain this phenomenon by the shock wave passage across the atmosphere. According to \cite {chadid}, there are important irregularities in the atmosphere of \object{RR~Lyr} that occur during the rising branch of successive pulsation cycles. This effect can also be responsible for the widening of the rising branch in the pg1 data set of \object{RR~Gem}. The irregular behaviour on the rising branch of \object{RR~Lyr} pointed out by \cite{cc} is, however, connected with the modulation. Therefore, as steady period change can be excluded, the larger scatter on the rising branch of \object{RR~Gem} indicates the presence of modulation.

\begin{figure}[h]
  \centering
  \includegraphics[height=7.5cm]{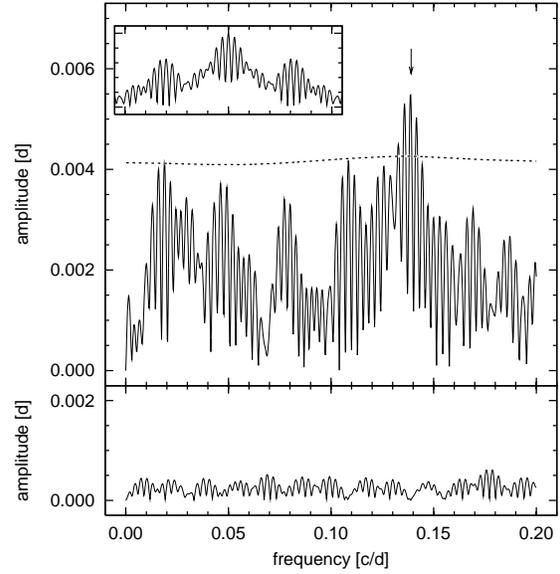}
  \caption{Fourier amplitude spectrum of the $(O-C)_{\mathrm{m}=11.3}$ data of the pg1 observations. The insert shows the spectral window function. The dotted line shows the $4\,\sigma$ significance level. An arrow points to the peak at $f=0.1389\,\mathrm{c/d}$. Bottom panel shows the residual spectrum after prewhitening with this frequency.}
  \label{fig_pg1_rb_oc_sp}
\end{figure}

\begin{figure*}[h]
  \centering
  \includegraphics[height=12cm,width=18cm]{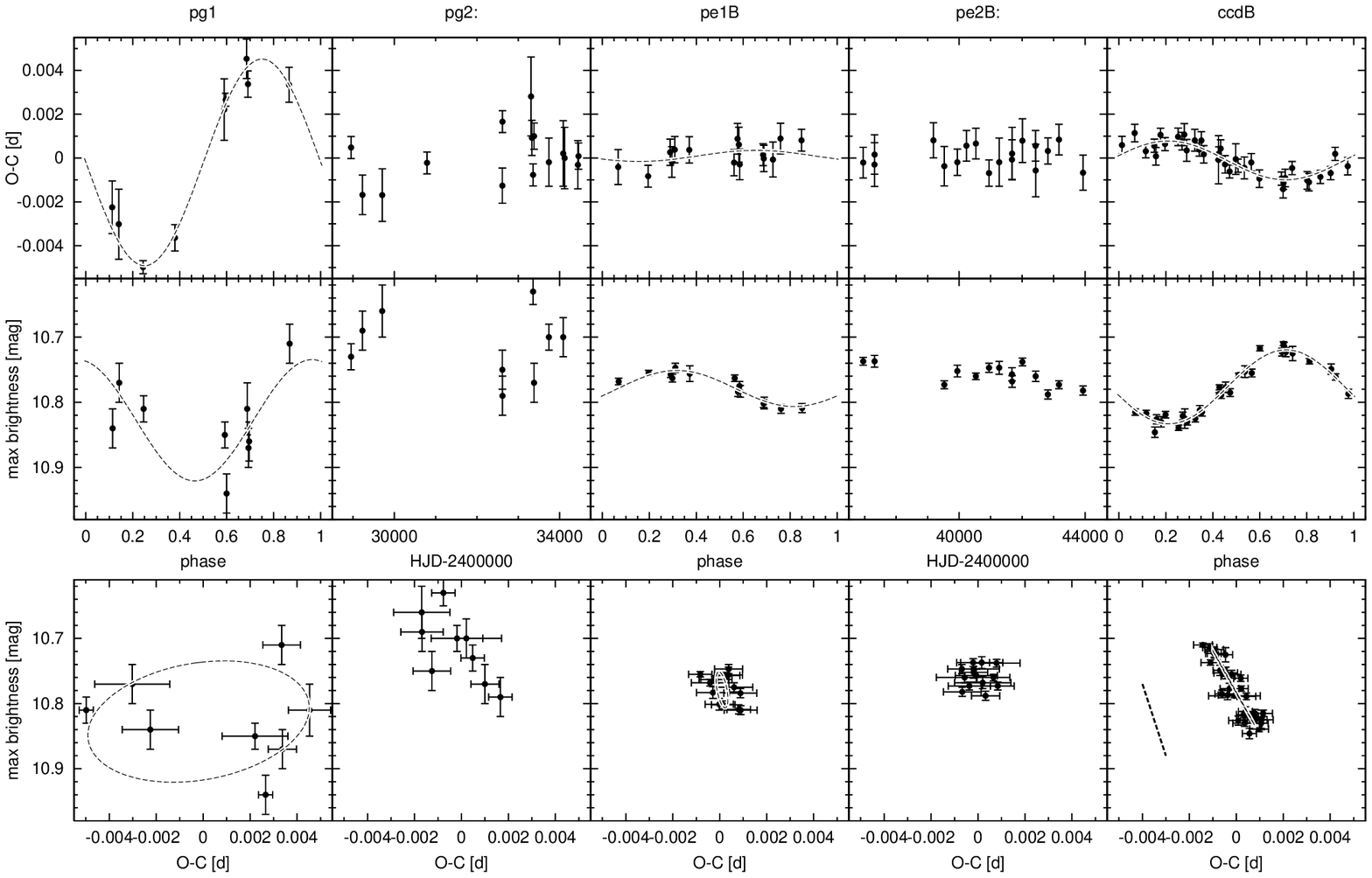}
  \caption{
The modulation in $(O-C)_{\mathrm{m}=11.3}$ (top), in maximum brightness (middle) and the relation between them (bottom) for the studied light curves. The $(O-C)_{\mathrm{m}=11.3}$ and maximum brightness points are folded with the determined modulation period for the pg1, pe1, and CCD data, and are plotted versus HJD for the pg2 and pe2 data. The fitted harmonic curves are also shown. In the second bottom panels, each point represents a particular pulsation cycle for which both $(O-C)_{\mathrm{m}=11.3}$ and maximum brightness could be determined. These plots are independent of the modulation period and reflect the amplitude relation and phase connection of the amplitude and phase modulation components. The plotted ellipses are the combinations of the fitted harmonic curves. In the bottom right panel we show the relation valid for test data of exact amplitude modulation.
  }
  \label{fig_tojas}
\end{figure*}

The modulation of the pg1 interval is first investigated in the rising branch phase data. These data, calculated from the $t_{\mathrm{m}=11.3}$ times (see electronic Table~13) according to the ephemeris in Sect.~\ref{sect_pglc}, were Fourier-analysed. The plausible modulation frequency range of $0-0.2\ \mathrm{c/d}$ were examined. The Fourier amplitude spectrum of the pg1 $(O-C)_{\mathrm{m}=11.3}$ data is plotted in the top panel of Fig.~\ref{fig_pg1_rb_oc_sp}. The most significant peak of the spectrum is at 0.139\,c/d frequency. Prewhitening the data with this frequency yields the residual spectrum shown in the bottom panel of Fig.~\ref{fig_pg1_rb_oc_sp}. This spectrum only shows low residual noise. The best sine curve fit to the $(O-C)_{\mathrm{m}=11.3}$ data gives \hbox{$f_{\mathrm{mod\,pg1}}=0.1389\pm0.0004\,\mathrm{c/d}$},  \hbox{$P_{\mathrm{mod\,pg1}}=7.20\pm0.02\,\mathrm{d}$}, and \hbox{$A_{\mathrm{phmod\,pg1}}=0.0094\pm0.0004\,\mathrm{d}$}. $A$ denotes peak-to-peak amplitude throughout this paper. The \hbox{$(O-C)_{\mathrm{m}=11.3}$} data phased with this period and the fitted harmonic curve are shown in the top left panel of Fig.~\ref{fig_tojas}. The $1^\mathrm{st}$ order harmonic curve fit the data with very small scatter, giving convincing evidence of phase modulation.

Due to the large scatter and significant nightly systematic errors, the maximum brightness magnitudes from the pg1 interval show the modulation period with uncertainty. These data cannot be used to refine the modulation period valid for this interval, but by accepting the period of the modulation from the $(O-C)_{\mathrm{m}=11.3}$ data, we can estimate the extent of the amplitude modulation. The middle left panel of Fig.~\ref{fig_tojas} shows the maximum brightness data folded with the 7.20\,d modulation period. A sine curve fit with this period gives \hbox{$A_{\mathrm{ampmod\,pg1}} = 0.19 \pm 0.07\,\mathrm{mag}$} and $\overline{B}_\mathrm{pg1\,max} = 10.83\pm0.02$\,mag values for the modulation amplitude and the average maximum brightness, respectively. The fact that the maximum brightness values can be phased with the period derived from the $(O-C)_{\mathrm{m}=11.3}$ data strengthens the finding of modulation during this interval.

\subsubsection{Modulation properties of the Pg2 data (JD\,$2\,428\,953-2\,434\,457$)}

The unfavourable data distribution makes the Fourier amplitude spectra of the pg2 maximum brightness and $(O-C)_{\mathrm{m}=11.3}$ data uninterpretable because of the severe aliasing effects and the possible change in the modulation period.

Though the period of the modulation cannot be derived from the pg2 data, we are able to estimate the amplitudes of the phase and amplitude modulations. In the bottom panels of Fig.~\ref{fig_tojas} the measured $(O-C)_{\mathrm{m}=11.3}$ versus maximum brightness data of the individual pulsation cycles are plotted. These plots are independent of the actual value of the modulation period. Any correlation or connection between these data is the sign that the measured scatters of the $(O-C)_{\mathrm{m}=11.3}$ and maximum brightness values are due to modulation rather than observational inaccuracy or any irregular behaviour. The bottom panel of the pg2 data in Fig.~\ref{fig_tojas} shows that this is indeed the case, since the variations in $(O-C)_{\mathrm{m}=11.3}$ and in maximum brightness are not independent. The possible extent of the amplitudes of the modulation can also be read from this plot: $A_{\mathrm{ampmod\,pg2}\,B}=0.14\,\mathrm{mag}$ and $A_{\mathrm{phmod\,pg2}\,B}=0.003\,\mathrm{d}$. If we assume that the pg1 and pg2 data have similar noise statistics and that the ratio of the modulation amplitudes and the scatters of the data are identical for the two data sets, then we can estimate the modulation amplitudes of the pg2 data simply from the observed scatter of the  maximum brightness and $(O-C)_{\mathrm{m}=11.3}$ values (see Table~\ref{tbl_maxok}). The modulation amplitudes estimated in this way are in very good agreement with those read from Fig.~\ref{fig_tojas}, confirming that the derived amplitudes of the modulation are reliable.

The average maximum brightness was $10.71 \pm 0.05$\,mag according to the pg2 data, which was $0.12 \pm 0.07$\,mag brighter than observed during the pg1 period. Although the scatter around minimum light made the estimation of the mean pulsation amplitude somewhat uncertain, it had increased with about $0.15-0.25$\,mag from the pg1 to the pg2 interval.

To check whether the observed change of the mean pulsation amplitude and/or the maximum brightness is real or is only an instrumental effect, we compared the mean photographic magnitudes of each of the comparisons stars separately for the pg1 and pg2 data. The magnitudes show no systematic variations between the two intervals, and the magnitude differences are less than 0.04\,mag for each star. Consequently, we regard the observed 0.14 mag increase in the average maximum brightness as real, and it is most probably a sign of real change in the mean pulsation amplitude as well.

\subsection{Results from photoelectric observations}
\label{sect_pelc}

The photoelectric observations made without any filter are quite inhomogeneous as they were obtained during the test phase of the new photometer with several different instrument settings. Therefore, this data set is only suitable for determining the $t_\mathrm{max}$ data but not for investigating the modulation.

The $t_\mathrm{max}$ data of the unfiltered observations, the times, and magnitudes of all the observed $B$ and $V$ light maxima, as well as the $t_{\mathrm{m}=11.3}$ timing data of the photoelectric light curves, are listed in Table~14.

The pulsation period changed with a constant rate between JD\,2\,435\,000 and JD\,2\,445\,000. The change rate has increased with about an order of magnitude in the years after JD\,2\,445\,000. The time distribution of the observations are uneven as nearly the half of the data points were obtained during the first 2 seasons (pe1 data, before JD\,2\,436\,629). Later, the observations were very sparse: $1-2$ nights per annum (pe2 data). The pe2 data contain the observations made between JD\,2\,536\,956 and JD\,2\,543\,931, as the period change rate increased later.

The period of the pulsation during the pe1 and pe2 observations was
$$P_\mathrm{pe} = P_\mathrm{pe\,0} + \dot{P}_\mathrm{pe} \cdot (t-t_\mathrm{pe\,0})\,\mathrm{d},$$
where $$P_\mathrm{pe\,0} = 0.3973148 \pm 2 \cdot 10^{-7}\,\mathrm{d},$$
$$\dot{P}_\mathrm{pe\,0} = -(7.4 \pm 0.2) \cdot 10^{-10}\,\mathrm{d/d},\ \mathrm{and}\ t_\mathrm{pe\,0} = 2\,536\,229.0\,\mathrm{[HJD]}.$$
The period and its change rate were determined by a $2^\mathrm{nd}$ order polynomial fit to the Konkoly photoelectric $(O-C)_{\mathrm{m}=11.3}$ data.

The $(O-C)_{\mathrm{m}=11.3}$ values were calculated from the pe1 and pe2 $t_{\mathrm{m}=11.3}$ times (listed in Table~14) according to the following ephemeris:
$$ t_{\mathrm{m=11.3\,pe}} = 2\,436\,229.0727\,{\rm [HJD]} + P_{\mathrm{pe\,0}} \cdot E_{\mathrm{pe}} - 1.47 \cdot 10^{-10} \cdot E_{\mathrm{pe}}^2. $$

\begin{figure}[h]
  \centering
  \includegraphics[height=4.5cm,width=7cm]{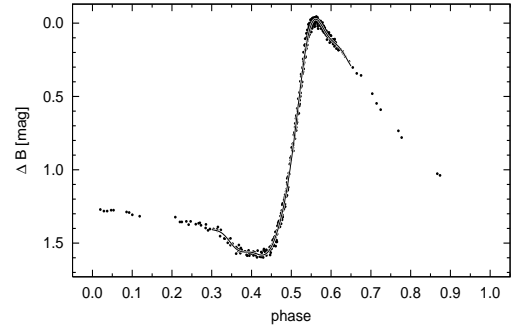}
  \caption{The pe1 photoelectric $B$ observations phased with the pulsation period and the fitted mean light curve.}
  \label{fig_pe_lc}
\end{figure}

The pe1 $B$ data folded with the average pe1 pulsation period and the fitted mean light curve taking 13 harmonics into account are plotted in Fig.~\ref{fig_pe_lc}.

\subsubsection{Modulation properties of the Pe1 data (JD\,$2\,436\,229-2\,436\,628$)}

\begin{figure}[h]
  \centering
  \includegraphics[width=8.6cm]{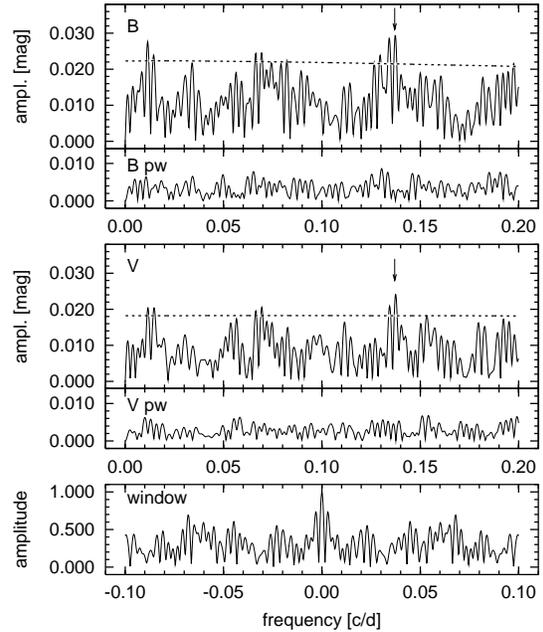}
  \caption{Fourier amplitude spectra of the pe1 $B$ and $V$ maximum brightness data. Dotted lines show the $4\,\sigma$ significance level. Arrows point to the peaks at $f=0.1374\,\mathrm{c/d}$. Panels marked with `pw' show the residual spectra after prewhitening with this frequency. Bottom panel shows the spectral window function.}
  \label{fig_pe_ampmod_sp}
\end{figure}

\begin{figure}[h]
  \centering
  \includegraphics[height=6.5cm,width=8.2cm]{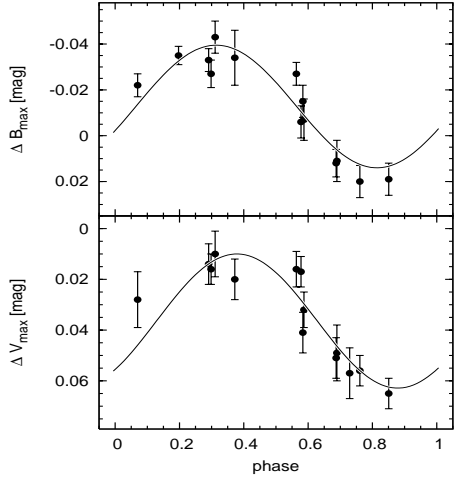}
  \caption{Photoelectric $B$ and $V$ maximum brightnesses of the pe1 data folded with the 7.28\,d modulation period.}
  \label{fig_pe_ampmod}
\end{figure}

The light curve in Fig.~\ref{fig_pe_lc} shows no sign of phase modulation on the rising branch; however, the scatter in magnitude around the light maximum is higher than the uncertainty of the observations that indicates variation in amplitude, either in a regular or an irregular way. The effect of period change is negligible during the pe1 interval. The Fourier amplitude spectra of the pe1 $B$ and $V$ maximum brightness data are shown in Fig.~\ref{fig_pe_ampmod_sp}. Both spectra show two peaks around the expected 0.139\,c/d frequency value. The frequency that fits both the $B$ and $V$ data sets the best is $f_\mathrm{mod\,pe1} = 0.1374 \pm 0.0003\,\mathrm{c/d}$ (i.e. $P_\mathrm{mod\,pe1} = 7.28 \pm 0.02\,\mathrm{d}$). The distance between this peak and its neighbour is 0.003\,c/d. The $-1\,$cycle/year alias frequency results in a fit with significantly higher rms. The pe1 $\Delta B$ and $\Delta V$ maximum brightness data phased with $P_\mathrm{mod\,pe1} = 7.28\,\mathrm d$ are plotted in Fig.~\ref{fig_pe_ampmod}. The amplitudes of the fitted harmonic curves are \hbox{$A_{\mathrm{ampmod\,pe1}\,B}=0.054\pm0.006\,\mathrm{mag}$} and \hbox{$A_{\mathrm{ampmod\,pe1}\,V}=0.052\pm0.007\,\mathrm{mag}$} for the $B$ and $V$ data, respectively. The average maximum brightness in $B$ is $10.77\pm0.01\,\mathrm{mag}$. According to the fitted mean light curve, the mean pulsation amplitude in $B$ band is $1.61\pm0.02\,\mathrm{mag}$.

\subsubsection{Modulation properties of the Pe2 data (JD\,$2\,436\,956-2\,445\,673$)}

The sparse data distribution in time and the pulsation period change during this interval makes these data hard to interpret. Therefore, the pe2 data are treated similarly to those of pg2. Upper limits of the amplitudes of the possible modulation are estimated from the plots shown in Fig.~\ref{fig_tojas} and from the comparison of the parameters of the pe1 and pe2 data listed in Table~\ref{tbl_maxok}. The estimations were made assuming identical \hbox{$\sigma/A$} ratios for the two photographic data sets and also for the two photoelectric data sets.

The scatter of the maximum brightness magnitudes of the pe2 data is more than 3 times greater than the average error of the individual points. It suggests that modulation at this time might also exist, but with even slightly smaller amplitude than in the first part of the photoelectric observations. The $(O-C)_{\mathrm{m}=11.3}$ data show no sign of phase modulation. The average maximum brightness of the pe2 $B$ light curve was $10.76 \pm 0.02\,\mathrm{mag}$ and the mean pulsation amplitude was the same as for the pe1 $B$ data within the error range.

\begin{table}[h]
  \centering
  \caption{Measured scatter of the maximum brightness and \hbox{$(O-C)_{\mathrm{m}=11.3}$} data  and the measured and estimated (italics) amplitudes of their variations.}
  \label{tbl_maxok}
  \begin{tabular}{lcc|cc}
    \hline\hline
    \noalign{\smallskip}
    data set & $\sigma(B_\mathrm{max})$ & $A_\mathrm{ampmod}$ &
               $\sigma(O-C)_{\mathrm{m}=11.3}$ & $A_\mathrm{phmod}$\\

             & [mag]   & [mag] & [d]  & [d] \\
    \noalign{\smallskip}
    \hline
    \noalign{\smallskip}
    pg1     & 0.065 &     0.19 & 0.0034 & 0.0094 \\
    pg2     & 0.052 & \it 0.15 & 0.0011 & \it 0.0030 \\
    \noalign{\smallskip}
    \hline
    \noalign{\smallskip}
    pe1 $B$ & 0.022 &     0.05 & 0.0005 & 0.0005 \\
    pe2 $B$ & 0.017 & \it 0.04 & 0.0004 & \it 0.0004 \\
    \noalign{\smallskip}
    \hline
  \end{tabular}
\end{table}

\subsection{Results from CCD observations}

A detailed study of the modulation properties of RR Gem during the 2004 observations was given in Paper1. Here we only summarise the global parameters of the modulation in order to compare them with those observed during the photographic and photoelectric observations. The $t_{\mathrm{m}=11.3}$ and maximum brightness data determined for CCD $B$ light curves of the 2004 and 2005 seasons are listed in Table~15. The $(O-C)_{\mathrm{m}=11.3}$ values were calculated according to the ephemeris:
$$ t_{\mathrm{m=11.3\,CCD}} = 2453019.5404\,{\rm [HJD]} + P_\mathrm{CCD} \cdot E, $$
where the pulsation period is $P_\mathrm{CCD} = 0.3972893\pm 3 \cdot 10^{-7}\,\mathrm{d}.$

The $(O-C)_{\mathrm{m}=11.3}$ and maximum brightness data of $B$ light curves are shown in the right panels of Fig.~\ref{fig_tojas}. These plots were constructed using the modulation period valid for the two seasons: \hbox{$P_\mathrm{mod\,CCD} = 7.216 \pm 0.003\,\mathrm{d}$} \citep{padeu}. The amplitude of the modulation in $(O-C)_{\mathrm{m}=11.3}$ and maximum brightness are \hbox{$A_{\mathrm{phmod\,CCD}\,B}=0.0018 \pm 0.0002\,\mathrm d$} and \hbox{$A_{\mathrm{ampmod\,CCD}\,B} = 0.114 \pm 0.004\,\mathrm{mag}$}, respectively (see right panels of Fig.~\ref{fig_tojas}). The average maximum $B$ brightness of the CCD data was $10.76 \pm 0.01\,\mathrm{mag}$, and the mean pulsation amplitude was $1.62 \pm 0.01\,\mathrm{mag}$.

\section{Discussion of long-term changes}

The long-term changes in the pulsation and modulation of \object{RR~Gem} can be followed in Figs.~\ref{fig_tojas} and \ref{fig_longterm}. The properties of the modulation and pulsation at the studied intervals are summarised in Table~\ref{tbl_sum}. 

Each of the measured parameters (pulsation and modulation periods and amplitudes, phase difference between maximum brightness and $(O-C)_{\mathrm{m}=11.3}$ data, and mean maximum brightness) show much larger changes than their uncertainties allow.

\subsection{Changes in the modulation properties}
\label{sect_nature}

In Fig.~\ref{fig_tojas} the modulation in $(O-C)_{\mathrm{m}=11.3}$ and in maximum brightness, as well as the relation between them, are plotted for all the studied intervals. The top and middle panels show the $(O-C)_{\mathrm{m}=11.3}$ and maximum brightness values folded with the derived modulation periods of the pg1, pe1, and CCD data and the fitted harmonic curves. Because the modulation period of the pg2 and pe2 data could not be determined due to imperfect data sampling, these data are plotted versus the HJD of the measurements. In the bottom panels each point represents a particular pulsation cycle for which both $t_{\mathrm{m}=11.3}$ and maximum brightness could be determined reliably. These plots are independent of the period of the modulation, and reflect the amplitude relation and phase connection of the amplitude and phase modulation components. The plotted ellipses are the combinations of the fitted sine curves of the maximum brightness and $(O-C)_{\mathrm{m}=11.3}$ data for the pe1, pg1, and CCD data.

It is not easy to make an exact distinction between amplitude and phase modulation of RR Lyrae stars, because the shape of the light curve (i.e. the relative amplitudes and phase differences of the Fourier components) always change more or less during the modulation cycle. Even if a node on the rising branch exists, a slight oscillation in the times of maxima could happen (this is only the case of \object{RR~Gem} at the CCD observations; see Fig.~5 in \citealt{sscnc}). Especially problematic is how to correctly measure phase modulation. As described in Sect.~\ref{sect_ocmax}, the $(O-C)_{\mathrm{m}=11.3}$ variation measures a combination of the phase and amplitude modulations. For comparison, the bottom panel of the CCD data in Fig.~\ref{fig_tojas} line shows the $(O-C)_{\mathrm{m}=11.3}\,-$\,maximum brightness connection predicted from test data of the exact amplitude modulation (see Sect.~\ref{sect_ocmax}). The actual slope of the observed $(O-C)_{\mathrm{m}=11.3}$ vs. maximum-brightness plots of the larger amplitude modulations are flatter than for the test data. During the photoelectric observations, when the amplitude modulation was the smallest, if there was any, no modulation in $(O-C)_{\mathrm{m}=11.3}$ was detected.

The bottom panels in Fig.~\ref{fig_tojas} show that significant changes in the modulation behaviour of \object{RR~Gem} occurred during the past 70 years. The most significant change in the nature of the modulation happened at around JD\,2\,428\,800 in conjunction with the sudden change in the pulsation period. Prior to this event, a pronounced phase modulation was accompanied by the modulation in amplitude. After the sudden period increase the character of the modulation had changed. The phase of the middle of the rising branch show a small variation in excess, if any, to what is expected from exact amplitude modulation, while the amplitude of the amplitude modulation changes in the range of an order of magnitude. The bottom panels of Fig.~\ref{fig_tojas} show that, after the sudden pulsation period change, only the strength of the modulation shows significant changes, but its character mostly remains the same.

The parameters plotted in  Fig.~\ref{fig_longterm} show that the changes in the amplitudes of the $(O-C)_{\mathrm{m}=11.3}$ and maximum-brightness variations are the most tightly connected values. The correlated variations of these amplitudes point to the total `power' of the modulation changing with time. Only a few Blazhko stars have extended enough photometric data to follow the changes in the modulation properties. The best-studied case is \object{RR~Lyr} itself, where the amplitude of the modulation shows cyclic changes on a 4-year timescale \citep{szeidliauc}. In other Blazhko stars \object{XZ~Cyg}, \object{AR~Her} \citep{almar}, \object{AH~Cam} \citep{smith99}, \object{RW~Dra} \citep{balazsdetre}, there are definite observations showing that the character of their modulation can change considerably, although no detailed study comprising their entire photometric data has been published.

\subsection{Changes in the pulsation properties}
\label{sect_chinpp}

The most important and surprising finding of this study of long-term changes in \object{RR~Gem} is the definite change in the mean maximum brightness and pulsation amplitude in the middle of the photographic observations. We have detected a difference of about 0.12 mag between the mean maximum brightnesses of the pg1 and pg2 data, while the uncertainties allow only less than 0.05\,mag, according to the comparison stars photometry (see details in Sect.~\ref{sect_pgdata}).

It cannot be excluded that the changes in these parameters between the photographic and photoelectric intervals are also real, but because of the defects of the calibration of the $B_\mathrm{pg}$ magnitudes, they most probably arise from the differences between the photographic and photoelectric $B$ photometries.

The explanation of the Blazhko phenomenon suggested by \cite{dz} involves energy transfer from the fundamental radial to a nonradial mode. It predicts that the larger the amplitude of the modulation, the smaller the pulsation mean amplitude of Blazhko stars. When the modulation was the strongest (pg1 data), the pulsation amplitude of \object{RR~Gem} was about 0.1\,mag smaller than at any time later, which agrees with the model predictions of \cite{dz}.

There are very few long-term Blazhko star observations available that are homogeneous, extended, and accurate enough to investigate similar variations in the modulation and pulsation amplitudes. Extended long-term observations in the same band have been summarised and/or published for
\object{RS~Boo} \citep[and references therein]{nagya},
\object{XZ~Cyg} \citep[and references therein]{lacluyze},
\object{RW~Dra} \citep[and references therein]{cokon102},
\object{XZ~Dra} \citep[and references therein]{xzdra},
\object{AR~Her} \citep[and references therein]{almar},
\object{RR~Lyr} \citep[and references therein]{kolenberg,smith03, cokon99},
and \object{RV~UMa} \citep[and references therein]{rvuma}.
Nevertheless, only two of them (\object{RR~Lyr} and \object{AR~Her}) are appropriate for investigation of modulation amplitude and mean maximum-brightness changes.

\cite{almar} showed that in \object{AR~Her} the brightest maxima remained at the same brightness level \citep[see Fig.~12 in][]{almar}, although the amplitude of the modulation had changed considerably. This behaviour indicates that the mean brightness was also fainter in \object{AR~Her} when the modulation amplitude was large, similar to what we have found in \object{RR~Gem}. It supports the correlation of the modulation amplitude and the mean maximum brightness of Blazhko variables.

The behaviour of the modulation of \object{RR~Lyr} is, however, different. The mean maximum brightness seems to remain unaltered during the 4-year cycle, while the modulation amplitude changes considerably \citep[see Fig.~6 in][Fig.~2]{szeidliauc,preston}. Nonetheless, the cyclic behaviour of the variation in the modulation amplitude of this star also differs from the irregular changes of \object{RR~Gem} and \object{AR~Her}, therefore it can have a different physical origin.

The possible differences between the pg and pe data does not allow us to make any conclusions about further changes in the mean maximum brightness and pulsation amplitude values between these observations. During the pe and CCD observations, though the amplitude of the modulation has changed these properties of the pulsation remained the same, indicating that the above-mentioned correlation is not strictly valid.

\subsection{Connections between the pulsation and modulation properties}

\begin{figure}[h]
  \centering
  \includegraphics[height=4cm,width=8.2cm]{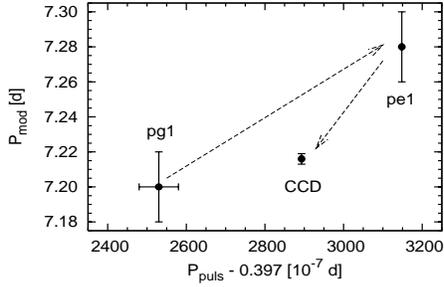}
  \caption{Modulation period versus pulsation period.}
  \label{fig_pppm}
\end{figure}

\begin{figure}[h]
  \centering
  \includegraphics[height=12cm,width=8.5cm]{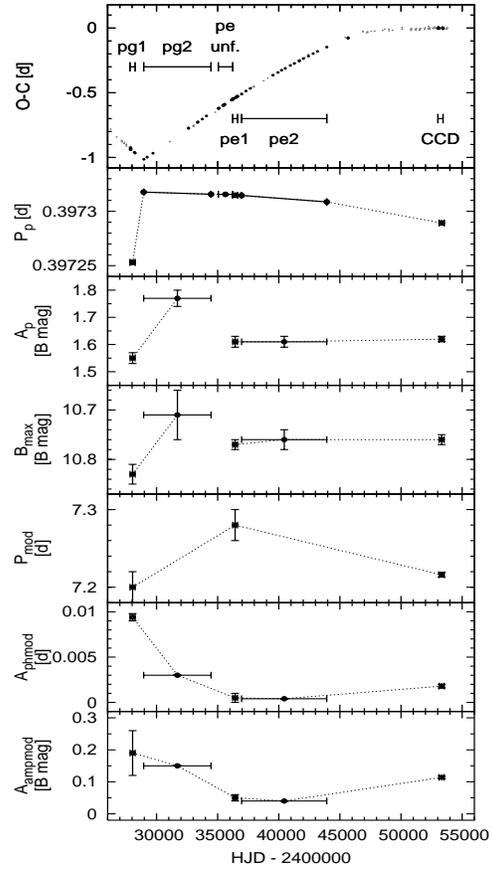}
  \caption{Variation of the pulsation and modulation properties of \object{RR~Gem}. The plotted quantities from top to bottom are $(O-C)_\mathrm{max}$, pulsation period, pulsation amplitude, average maximum brightness, modulation period, modulation amplitude in $(O-C)_{\mathrm{m}=11.3}$, and modulation amplitude in maximum brightness.}
  \label{fig_longterm}
\end{figure}

The period changes of the pulsation and modulation indicate a positive correlation between the periods  as shown in Fig.~\ref{fig_pppm}, although they can be compared only on the bases of three epochs of data. The data define $\mathrm d\,P_\mathrm{mod}/\mathrm d\,P_\mathrm{puls}=1.6\pm0.8\times10^3$ for \object{RR~Gem}, which completes the list of Blazhko stars with measured period change of both their periods  \cite[][Table~8]{lacluyze} with a new item.

It was shown in \cite{ibvs} that the largest possible amplitude of the modulation depends on the pulsation frequency of Blazhko stars. However, at a given pulsation frequency, the modulation has very different amplitudes, from very low values to the extent of the possible maximum value. That those Blazhko variables with long-term photometric observations show significant changes in their modulation amplitudes proves that the measured amplitude of the modulation is just a temporal parameter. It may explain why the pulsation frequency gives only a limit for the possibly highest value of the modulation amplitude and does not determine its actual value.

Most probably the `instability' of the modulation is  a general property of Blazhko stars, and any plausible explanation of the phenomenon should give a reason for the irregularity of the observed light curve modulation \citep[see also][]{cc}.

What the modulation properties (amplitude, phase relation between the amplitude, phase modulation, etc.) tell us about the physics of the phenomenon, if they show considerable changes on larger timescales, is unclear. During the time of the photoelectric observations, \object{RR~Gem} showed hardly any modulation. This fact warns us that the modulation can only be temporarily detected in some cases. In contrast to previous expectations, the amplitude of the modulation can also be small \citep{sscnc}, below the detectability limit of most of the extended surveys. Therefore, the most plausible conclusion is that the modulation is a common, intrinsic property of RR Lyrae stars.

\begin{table*}[]
  \centering{
  \caption{Modulation and pulsation properties derived from the observations.}
  \label{tbl_sum}
  \begin{tabular}{lllllllr}
    \hline\hline
    \noalign{\smallskip}
    \multicolumn{1}{c}{data set} & \multicolumn{1}{c}{$P_\mathrm{puls}$(error)} & \multicolumn{1}{c}{$P_\mathrm{mod}$(error)} &
    \multicolumn{1}{c}{$A_\mathrm{puls}$(error)} & \multicolumn{1}{c}{$\overline{B}_\mathrm{max}$(error)\,$^a$} & 
    \multicolumn{1}{c}{$A_\mathrm{phmod}$(error)} & \multicolumn{1}{c}{$A_\mathrm{ampmod}$(error)} & 
    \multicolumn{1}{c}{$\Delta$\,phase(error)\,$^b$} \\

    &          \multicolumn{1}{c}{[d]}             & \multicolumn{1}{c}{[d]}                     & \multicolumn{1}{c}{[mag]}
    & \multicolumn{1}{c}{[mag]}                     &
    \multicolumn{1}{c}{[d]}                     & \multicolumn{1}{c}{[mag]}                      &
    \multicolumn{1}{c}{[rad]}           \\
    \noalign{\smallskip}
    \hline
    \noalign{\smallskip}
pg1     & 0.397253(5)         & 7.20(2)  & 1.55(2) & 10.83(2) & 0.0094(4) & 0.19(7)  & -1.8(3)\\
pg2     & 0.397317(2)\,$^c$   &  -       & 1.77(3) & 10.71(5) & 0.003     & 0.15     & 0.6(7)\\
pe unf. & 0.3973156(35)       &  -       &  -      &   -      & -         & -        & -\\
pe1 $B$ & 0.3973148(2)        & 7.28(2)  & 1.61(2) & 10.77(1) & 0.0005(4) & 0.05(1)  & -\\
pe2 $B$ & 0.3973160(2)\,$^c$  &  -       & 1.61(2) & 10.76(2) & 0.0004    & 0.04     & -\\
CCD $B$ & 0.3972893(3)        & 7.216(3) & 1.62(1) & 10.76(1) & 0.0018(2) & 0.114(4) & 0.0(1)\\
    \noalign{\smallskip}
    \hline
    \noalign{\smallskip}
  \multicolumn{8}{p{153mm}}{$^a$ The errors of the mean maximum magnitudes were determined in two different ways. When the modulation period was found, it corresponds to the zero point uncertainty of the fitted sine curve. For the pg2 and pe2 data, the given errors are the scatters of the maximum points, which overestimate the true errors, if this scatter originates partially from modulation.}\\
  \multicolumn{8}{p{153mm}}{$^b$ $\Delta$\,phase denotes the difference between the phases of the maximum magnitude and $(O-C)_{\mathrm{m}=11.3}$ variation.}\\
  \multicolumn{8}{p{153mm}}{$^c$ As the pulsation period changed during the pg2 and pe2 observations, the given values correspond to the middle of these intervals. Details of the period change are given in Sects.~\ref{sect_pglc} and \ref{sect_pelc}.}
  \end{tabular}
  }
\end{table*}

\section{Conclusions}

The rise in atmospheric shock waves during the pulsation of RR~Lyrae stars is well-established both observationally and theoretically. \cite{preston} proposed that the depth of the formation of the main shock wave during rising light shifts periodically in the course of the Blazhko cycle. \cite{cc} report irregular changes in the atmosphere of \object{RR~Lyr} during the Blazhko cycle. They find that the residual scatter of the radial velocity curve is largest when the nonlinear effects are the most intense, which is during the shock wave passage across the atmosphere.

The uncertainty of our photographic data and the unfavourable data distribution of both the photographic and photoelectric data sets do not allow us to distinguish between observational errors and irregularities. However, the detected systematic variations give clear evidence of the light curve modulation. Analysis of the 70-year long photometric observations of \object{RR~Gem} has revealed some new and important properties of the Blazhko modulation.

\object{RR~Gem} is the first Blazhko RR~Lyrae star where the amplitude of the  pulsation showed definite changes during the observation times. An unambiguous amplitude increase happened together with a sudden pulsation period increase around JD\,2\,428\,800. Before this event, a clear and strong modulation was observable with an amplitude of nearly 0.2\,mag in maximum brightness and 14\,min in phase on the rising branch. After the sudden pulsation period change, the strength of the modulation decreased, along with an increase in the mean pulsation amplitude. The photoelectric observations of $1958-1959$ show clear but weak modulation only in amplitude, with a slightly longer Blazhko period. The $2004-2005$ CCD observations show modulation with an amplitude of about 0.1\,mag in $B$ band and with very weak phase modulation. 

Cyclic changes in the modulation and pulsation properties are known for two Blazhko variables, \object{RR~Lyr} \citep{szeidliauc} and \object{XZ~Dra} \citep{xzdra}. Supposing that the changes observed in RR Gem are also a part of a cyclic behaviour on a century-long timescale, both the pulsation and modulation periods are expected to decrease, while the amplitude of the modulation is expected to increase in the future. If a sudden jump in the pulsation and modulation properties of RR Gem, similar to that observed in 1937, would happen in the future that would give us a unique opportunity to study the connections of the pulsation and modulation properties of Blazhko variables in detail. The extremely short modulation period (7.2 d) makes it easier to check the modulation amplitude of RR Gem annually.

\begin{acknowledgements}

We would like to thank the referee, M. Chadid, for her useful suggestions on this paper. This research has made use of the SIMBAD database, operated at the CDS Strasbourg, France, and the GEOS RR~Lyrae database. The financial support of OTKA grants T-043504, T-046207, and T-048961 is acknowledged.

\end{acknowledgements}

\end{document}